\documentclass[conference]{IEEEtran}
\ifCLASSINFOpdf
 
\else
\fi

\usepackage{cite}
\usepackage{amsmath}
\usepackage{amssymb}
\usepackage{epsfig}
\usepackage{graphicx}
\usepackage{subfigure}

\begin{document}
%
\title{A Novel Compact Dual-Band Antenna Design for WLAN Applications}

\author{\IEEEauthorblockN{Peshal B. Nayak, Ramu Endluri, Sudhanshu Verma and Preetam Kumar }
\IEEEauthorblockA{Department of Electrical Engineering\\
Indian Institute of Technology Patna\\
Patna, Bihar 800013, India\\
Email: nayak.ee10@iitp.ac.in, endluri.ee10@iitp.ac.in, sverma@iitp.ac.in, pkumar@iitp.ac.in}
}

\maketitle

\begin{abstract}
A novel and compact dual band planar antenna for 2.4/5.2/5.8-GHz wireless local area network(WLAN) applications is proposed and studied in this paper. The antenna comprises of a T-shaped and a F-shaped element to generate two resonant modes for dual band operation. The two elements can independently control the operating frequencies of the two excited resonant modes. The T-element which is fed directly by a 50 $\Omega$ microstrip line generates a frequency band at around 5.2 GHz and the antenna parameters can be adjusted to generate a frequency band at 5.8 GHz as well, thus covering the two higher bands of WLAN systems individually. By couple-feeding the F-element through the T-element, a frequency band can be generated at 2.4 GHz to cover the lower band of WLAN system. Hence, the two elements together are very compact with a total area of only 11$\times$6.5 mm$^{2}$. A thorough parametric study of key dimensions in the design has been performed and the results obtained have been used to present a generalized design approach. Plots of the return loss and radiation pattern have been given and discussed in detail to show that the design is a very promising candidate for WLAN applications.\footnote{This document is an author's version of \cite{nayak2013compact}.}   
\end{abstract}


%
\IEEEpeerreviewmaketitle

\section{Introduction}
\IEEEPARstart{W}{ireless} communications have developed rapidly in the modern world and have covered many technical areas. Since FCC announced to allow potential users to make an unlicensed use of medical, industrial and scientific frequencies, the scientific community has seen a great opportunity to design wireless devices that would communicate over short distances. Common examples are the Bluetooth which operates at the internationally available ISM band at $2.4$ GHz or the wireless local area network (WLAN) which operates at $5.2$ and $5.8$ GHz. \\
\indent WLAN implemented as an alternative for or an extension to wired LAN is a flexible data-communications system. WLANs make use of radio frequency technology and transmit and receive data over the air. This minimizes the need for wired connections and combines connectivity with user mobility. Nowadays, WLANs are becoming popular in a number of vertical markets such as retail, healthcare, warehousing, manufacturing and academia which have profited from the use of hand-held terminals for real-time information transmission to centralized hosts for processing. WLANs are also being widely recognized as a reliable, cost effective solution for wireless high speed data connectivity and a general purpose connectivity alternative for a broad range of applications. There are three operation bands in the IEEE $802.11$ WLAN standards: $2.4$ GHz $(2400$-$2484$ MHz$)$, $5.2$ GHz $(5150$-$5350$ MHz$)$ and $5.8$ GHz $(5725$-$5825$ MHz$)$ \cite{song2008small}. WLANs working at IEEE $802.11$a employ the higher frequency band from $5.15$-$5.35$ GHz and $5.725$-$5.825$ GHz while those working at IEEE $802.11$b/g use the $2.4$-$2.484$ GHz band. $802.11$a is usually found on business networks due to its higher cost. \\
\indent Nowadays, dual band WLAN systems combining the IEEE $802.11$a/b/g standards are becoming more attractive \cite{cao2008compact, nayak2017multi, nayak2019ap, nayak2019modeling, nayak2019virtual, nayak2016performance, peshal2019modeling}. Hence, to satisfy the need of wireless communications, it is necessary to design compact high performance antennas with $2.4/5$ GHz dual band operation and excellent radiation characteristics. Wide impedance bandwidth, simple configuration, omnidirectional radiation pattern and low cost are some of the important features of a planar monopole antenna. Naturally, it is one of the most commonly used antenna models for WLAN systems.\\    
\indent A large number of WLAN antennas have been recently proposed and reported in literature. A technique of designing dual band antennas has been proposed in \cite{chang2009meandered, chu2010design, yeh2002dual, kim2005cpw, yildirim2006low, wu2006new, d2006printed, nayak2013multiband, nayak2014novel, nayak2012ultrawideband, endluri975low}. It suggests the use of one monopole for the lower band and another for the higher band of WLAN systems. However, this results in a large antenna size due to the large length of the monopole resonating in the lower band. Therefore, in order to achieve size reduction, an interesting method of bending this monopole to different shapes has been used in \cite{chang2009meandered, chu2010design, yeh2002dual}. With the use of FR4 substrate with a relative permittivity ($\epsilon_{r}$) of $4.4$, the smallest size achieved by using this technique was about $15\times10$ mm$^{2}$ \cite{yeh2002dual}. This is still larger than the $11\times6.5$ mm$^{2}$ size of our proposed antenna. Another effective size reduction technique is the use of an inverted-F structure \cite{azad2009miniature, gallo2011design, liu2005inverted, tie2006design}. However, for multiband operation, the use of additional radiating elements is essential. Thus, incorporating them with an inverted-F structure and achieving a compact size is a design challenge. This challenge has been tackled in \cite{razali2009coplanar} by using a direct fed compact sized inverted-F element and two other long slots on the ground plane to generate the $2.4$ GHz, $5.2$/$5.8$ GHz and $3.5$ GHz bands. However, the slots on the ground plane result in a larger overall size as compared to our compact design. The direct fed planar inverted-F antenna (PIFA), proposed in \cite{wang2011internal}, resonated in the fundamental mode at $2.4$ GHz and the second order mode at $5.2$ GHz. The PIFA was combined with a parasitic element which had one end shorted to the ground and was used to generate the $5.8$ GHz band. The antenna was high profile and occupied a larger volume than the planar antenna due to the PIFA structure, though its height had been significantly reduced.\\ 
\indent Another interesting design has been reported in \cite{su2010high}, whereby two loop antennas have been combined together to achieve a dual band operation. However, since both loops required operating in one wavelength resonant modes, the element was quite large. In \cite{cumhur2009dual}, the design generated a dual band operation by using a split-ring as a monopole radiator. However, the overall size, when fabricated on a substrate with a large $\epsilon_{r}$= $6.15$ was $16\times13$ mm$^{2}$. A slot, occupying a large area of $30\times14$ mm$^{2}$, in the ground plane was embedded with a pair of horizontal strips to achieve dual band performance in \cite{dang2010compact}. In order to obtain a concealed antenna for WLAN systems, that is an antenna without any protruded portions in its appearance, a printed double-T monopole antenna has been presented in \cite{liu2005wideband}. Though a good performance was achieved, the overall dimensions of the design was $51$(L)$\times75$(W)$\times0.8$(H) mm$^{3}$ which is very large as compared to our proposed antenna.
\indent In this paper, a very compact planar dual band monopole antenna to cover the WLAN operating bands of $2.4$ and $5.2$/$5.8$ GHz is proposed. A microstrip fed T-shaped element and an F-shaped element have been used to achieve a dual band operation. The F element has been coupled-fed through the T element by placing it very close to the T element. Thus, with the use of a single feed the overall size has been considerable reduced. The antenna is designed and studied with the high performance full-wave electromagnetic (EM) field simulator Ansoft HFSS software. The paper is organized as follows. In Section II, the details of the proposed antenna have been presented. This is followed by the results of a parametric study on the antenna in Section III. A generalized design methodology for achieving dual band operation at other operating frequencies has been given in Section IV. Results and discussion have been presented in Section V. Conclusion has been given in the last section.  

%
%
%
%

\section{Antenna Design}
\begin{figure}[!t]
\centering
\includegraphics[width= 3.8 in]{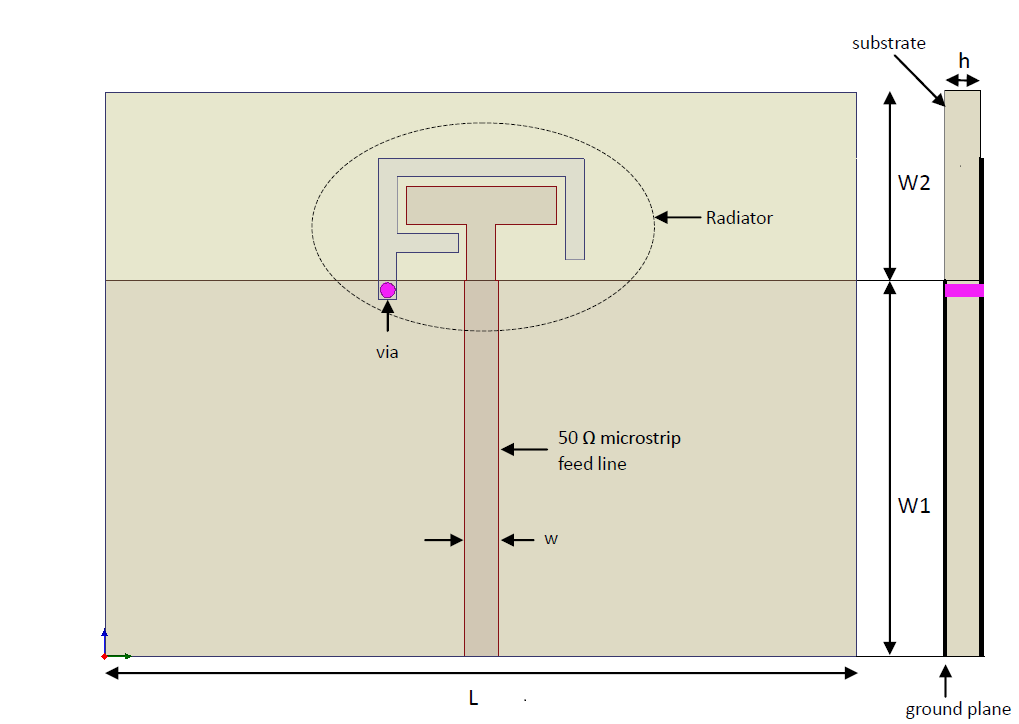}
\caption{Design of the dual band antenna}
\label{fig1}
\end{figure}

\begin{figure}[!t]
\centering
\includegraphics[width=3.8in]{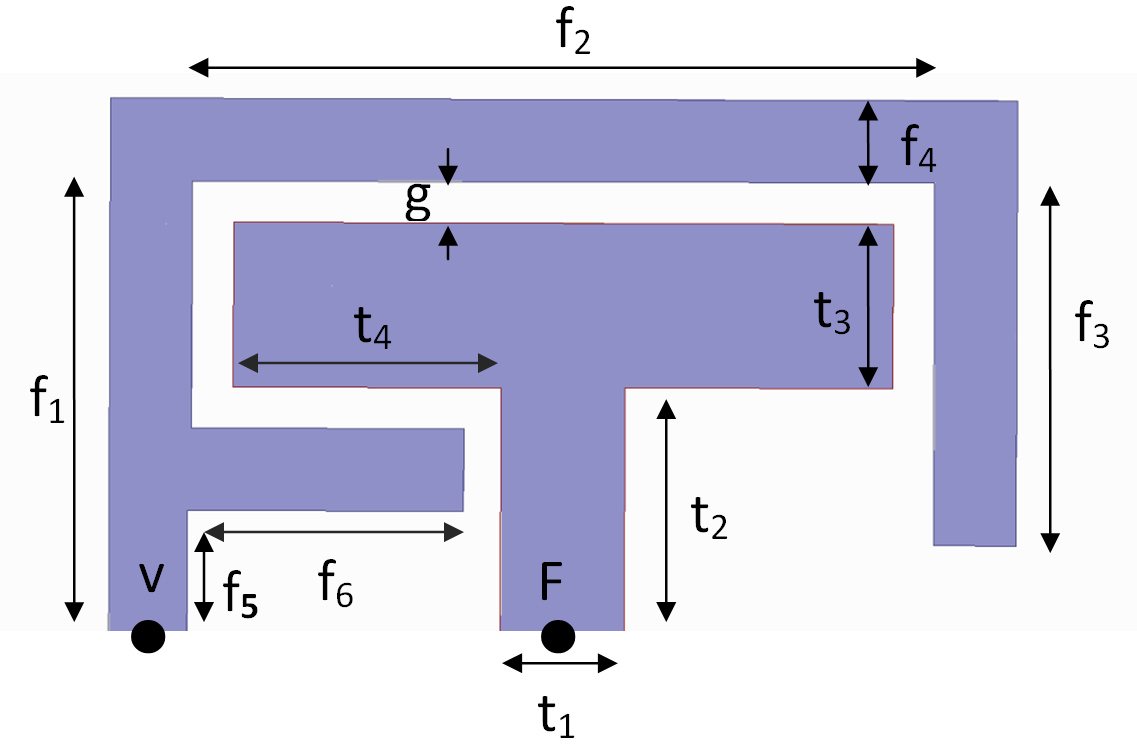}
\caption{Layout of radiator}
\label{fig2}
\end{figure}

The geometry of the proposed antenna has been shown in Fig.~\ref{fig1} and specifications of the design have been given in Table~\ref{table1}. A $50$ $\Omega$ microstrip feed line has been used in order to achieve a dual band operation with good impedance matching. The planar monopole antenna consists of a ground plane of $40$x$20$ mm$^{2}$ and a radiator of $11\times6.5$ mm$^{2}$ with overall dimensions of $40\times30\times0.8$ mm$^{3}$. Two radiating elements are present in the design. As these elements appear similar in shape to the letters T and F, they have been referred to as the T and F-elements. The compact size of the antenna results from the close packing achieved by the use of a single feed point for these two separate elements. The T- element has been fed by a microstrip feed line and is surrounded by a shorted parasitic F- element which has been placed coupled-fed from the T shape element via a small gap g. The diameter of the via used for shorting the F element to the ground is $0.4$ mm. The overall length, width and height of the antenna are represented by W, L and h respectively. For the T and F- element dimensions, the prefixes t and f respectively have been used in the layout in Fig.~\ref{fig2}. `F' is the feed point and `V' is the location of the via.\\
\indent The T element generates a band at $5.2$/$5.8$ GHz for the higher band of WLAN systems while the F element generates a band at $2.4$ GHz for the lower operating frequency of WLAN devices. As mentioned in Table I, the antenna has been fabricated on a substrate of relative permittivity $3.5$ and loss tangent $0.02$. Various dimensions of the antenna have been carefully optimized with the help of computer simulations. The antenna has been fabricated by using these optimized dimensions as shown in Table~\ref{table2}.

\begin{table}[t]
\caption{Antenna Design Details}
\centering
\begin{tabular}{|l|c|p{3 cm}|}
\hline
{\bf Material} &  FR$4$\\
\hline
{\bf Dielectric constant} & $3.5$\\
\hline
{\bf Loss Tangent} & $0.02$\\
\hline
{\bf Substrate Thickness} & $0.8$ mm\\
\hline
{\bf WLAN centre frequencies} & $2.4$ GHz, $5.2$ GHz, $5.8$ GHz\\
\hline
\end{tabular}
\label{table1}
\end{table}

\begin{table}[!]
\caption{Value of various dimensions in the antenna design (in mm)}
\centering
\begin{tabular}{|l|l|l|l|p{3 cm}|}
\hline
{\bf f1} &  $5.5$ & {\bf f2} & $9$\\
\hline
{\bf f3} & $4.4$ & {\bf f4} & $1$ \\
\hline
{\bf f5} & $2.5$ & {\bf f6} & $3.3$\\
\hline
{\bf t1} & $1.5$ & {\bf t2} & $3$\\
\hline
{\bf t3} & $2$ & {\bf t4} & $3.25$\\
\hline
{\bf g} & $0.5$ &  {\bf h} & $0.8$\\
\hline
{\bf L} & $40$ &  {\bf W1} & $20$\\
\hline
{\bf W2} & $10$ & {\bf w} & $1.8$\\
\hline

\end{tabular}
\label{table2}
\end{table}

\begin{figure}[!]
\centering
\includegraphics[width=1\linewidth]{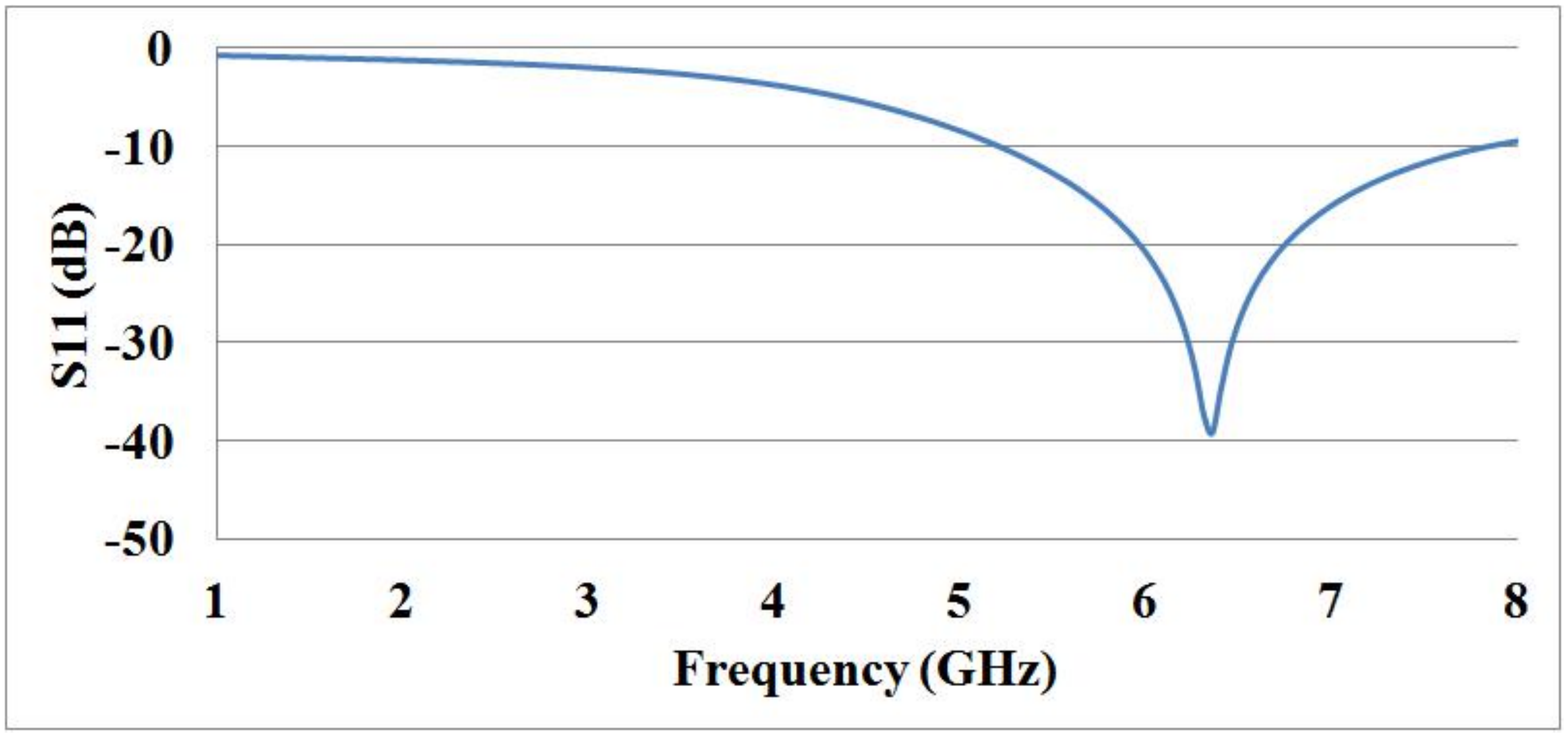}
\caption{Return Loss plot of the antenna design with only T-element present}
\label{fig5}
\end{figure}

\begin{figure}[!]
\centering
\includegraphics[width=0.9\linewidth]{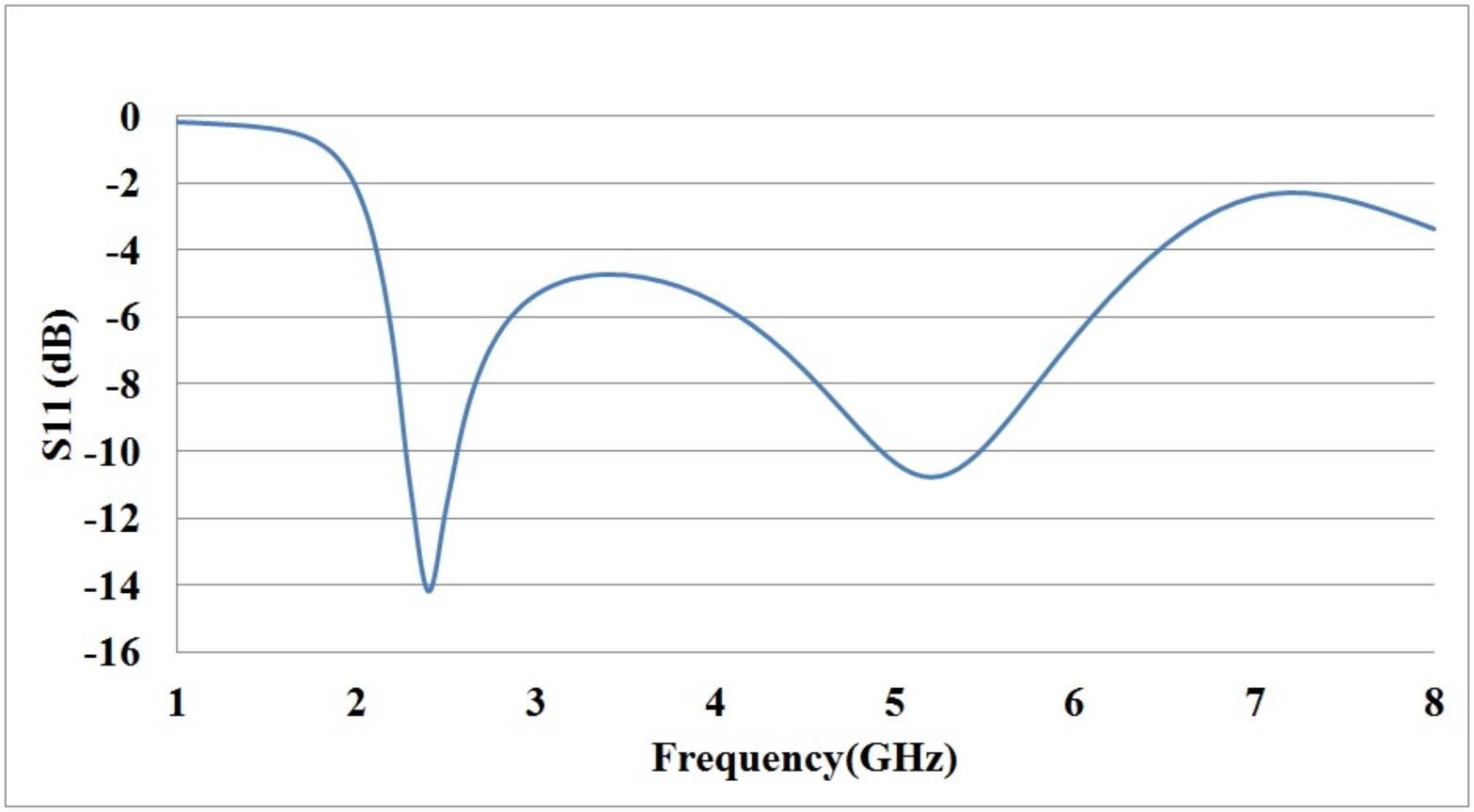}
\caption{Return Loss plot of the antenna design with optimized parameters (both elements present)}
\label{fig3}
\end{figure}

\section{Parametric Study}
\begin{figure}[!]
  \centering
  \subfigure[]
  {
      \includegraphics[width=1\linewidth]{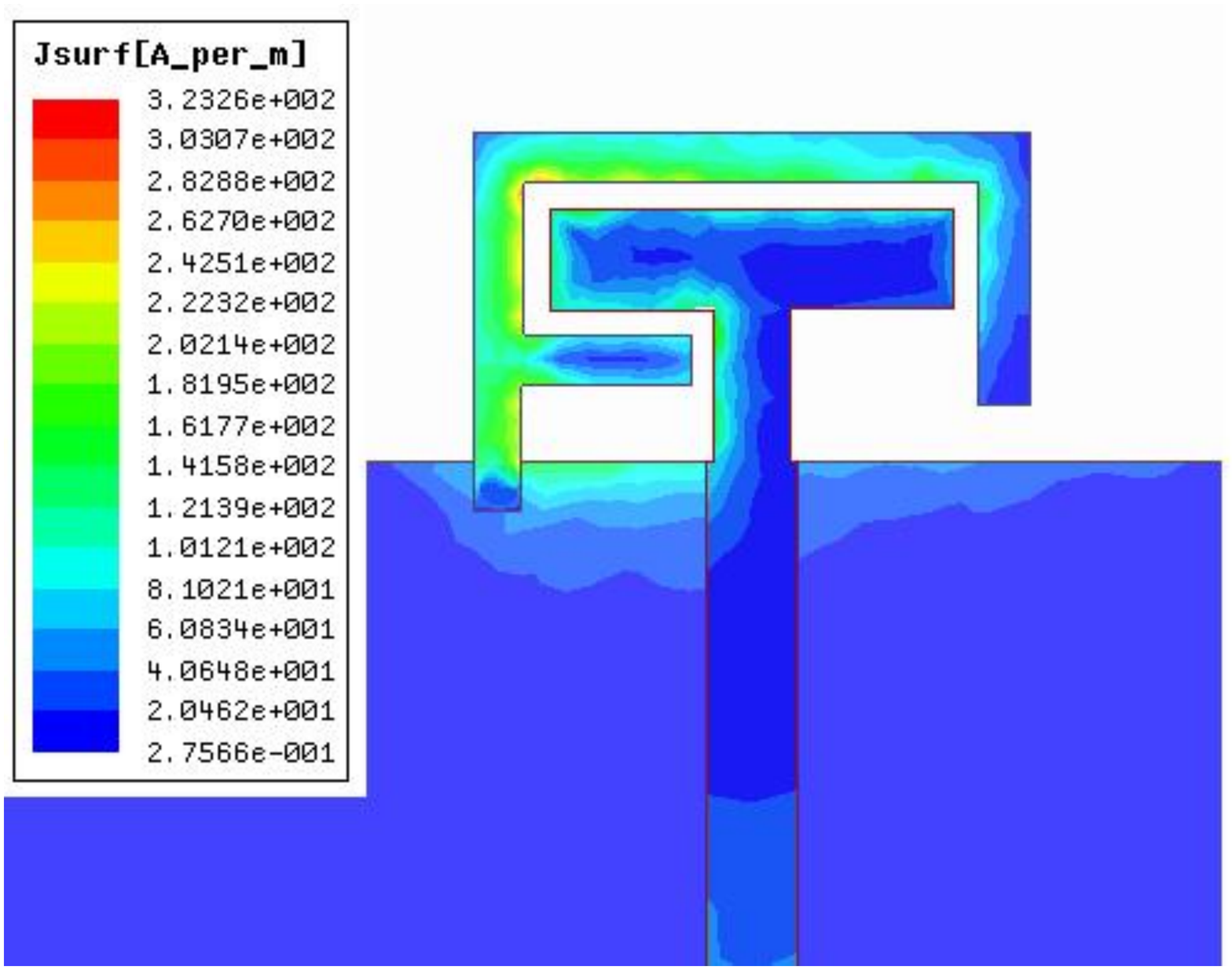}
      \label{ellipse}
  }
  \subfigure[]
  {
      \includegraphics[width=1\linewidth]{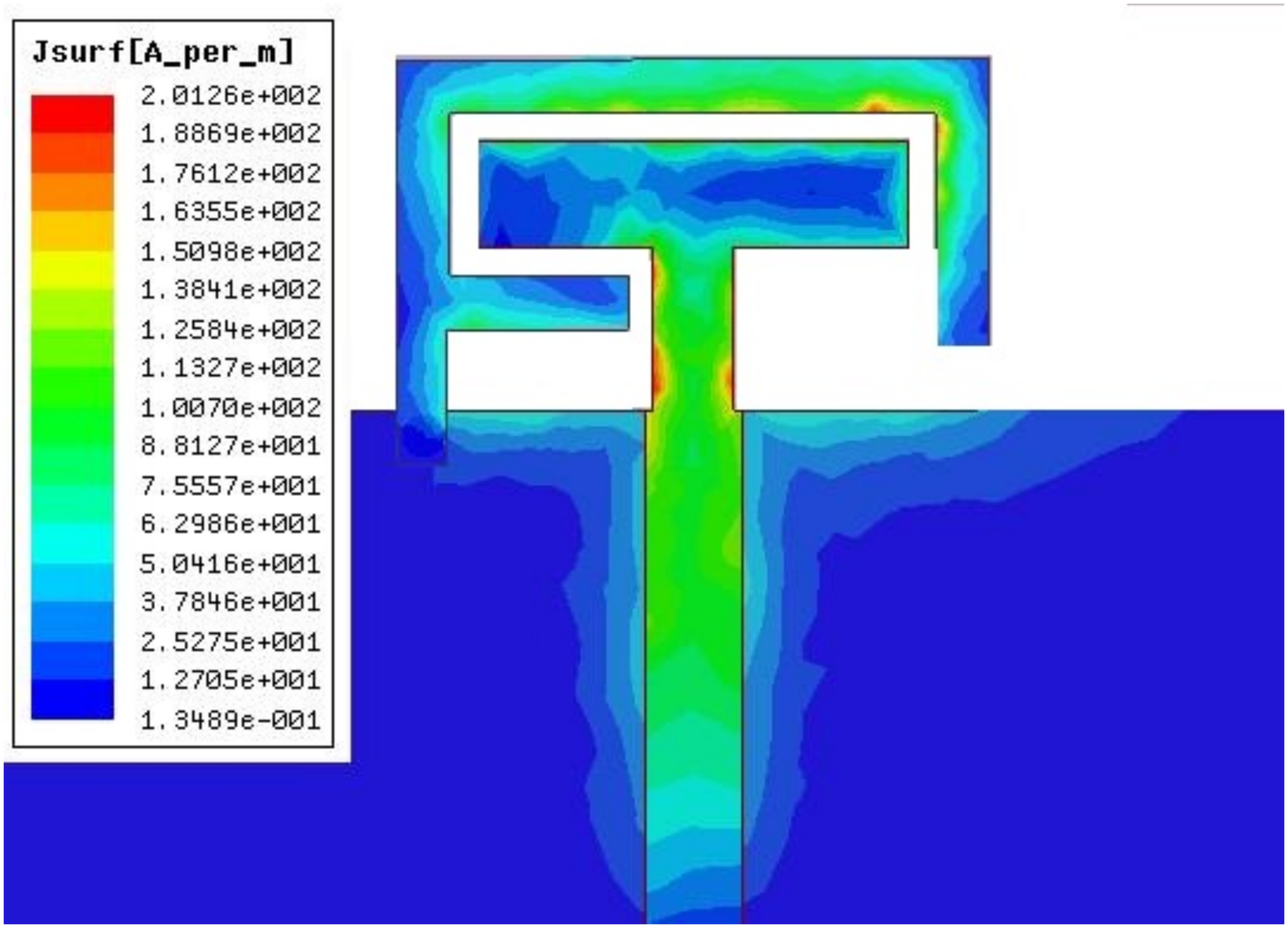}
      \label{circle}
  }
    
\caption{Simulated current distribution at (a)2.4 GHz (b)5.2 GHz}
  \label{fig6}
\end{figure}

\begin{figure}[t]
\centering
\includegraphics[width=1\linewidth]{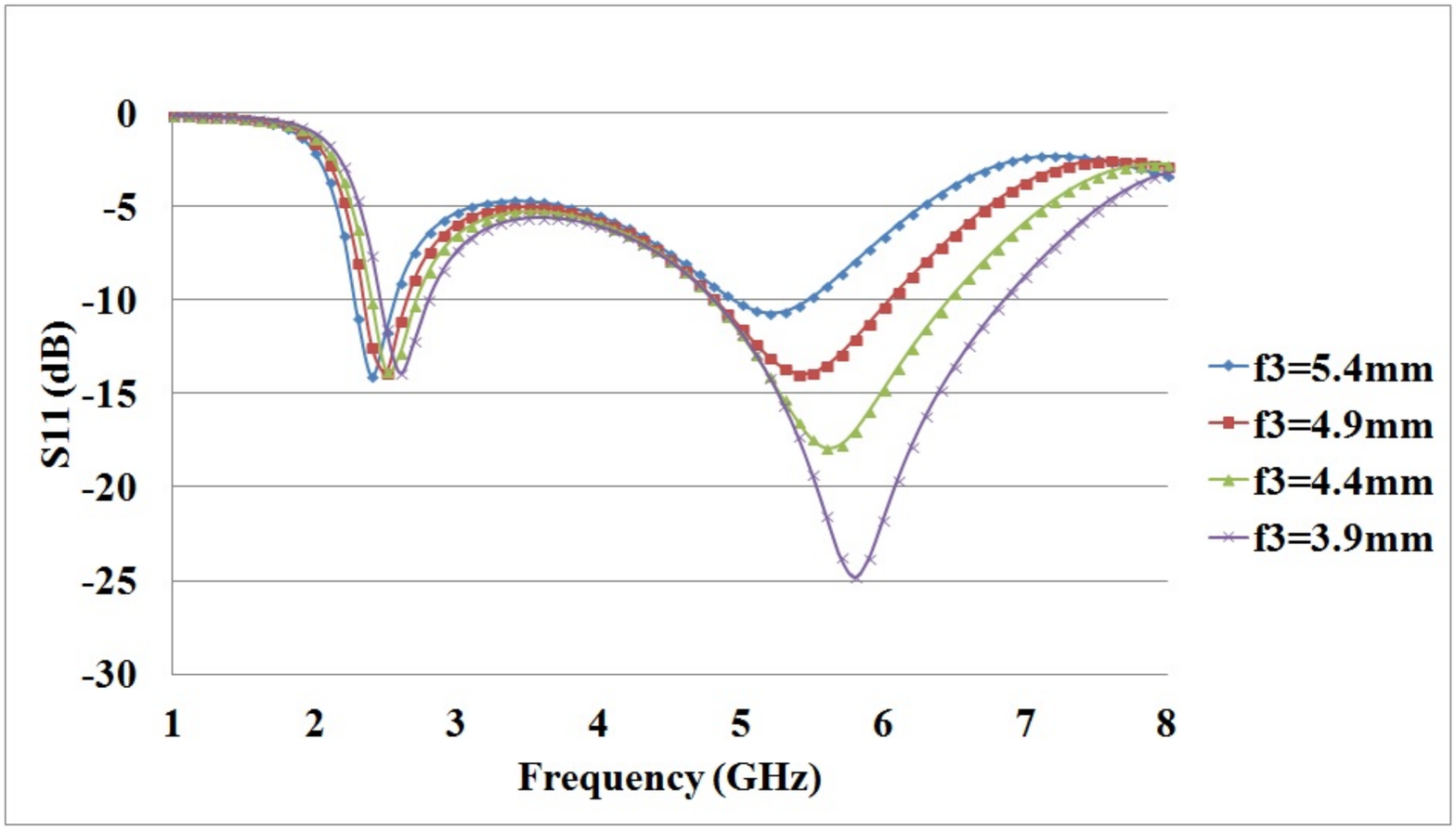}
\caption{Return Loss with variation in the length of f$3$}
\label{fig7}
\end{figure}

\begin{figure}[!]
\centering
\includegraphics[width=1\linewidth]{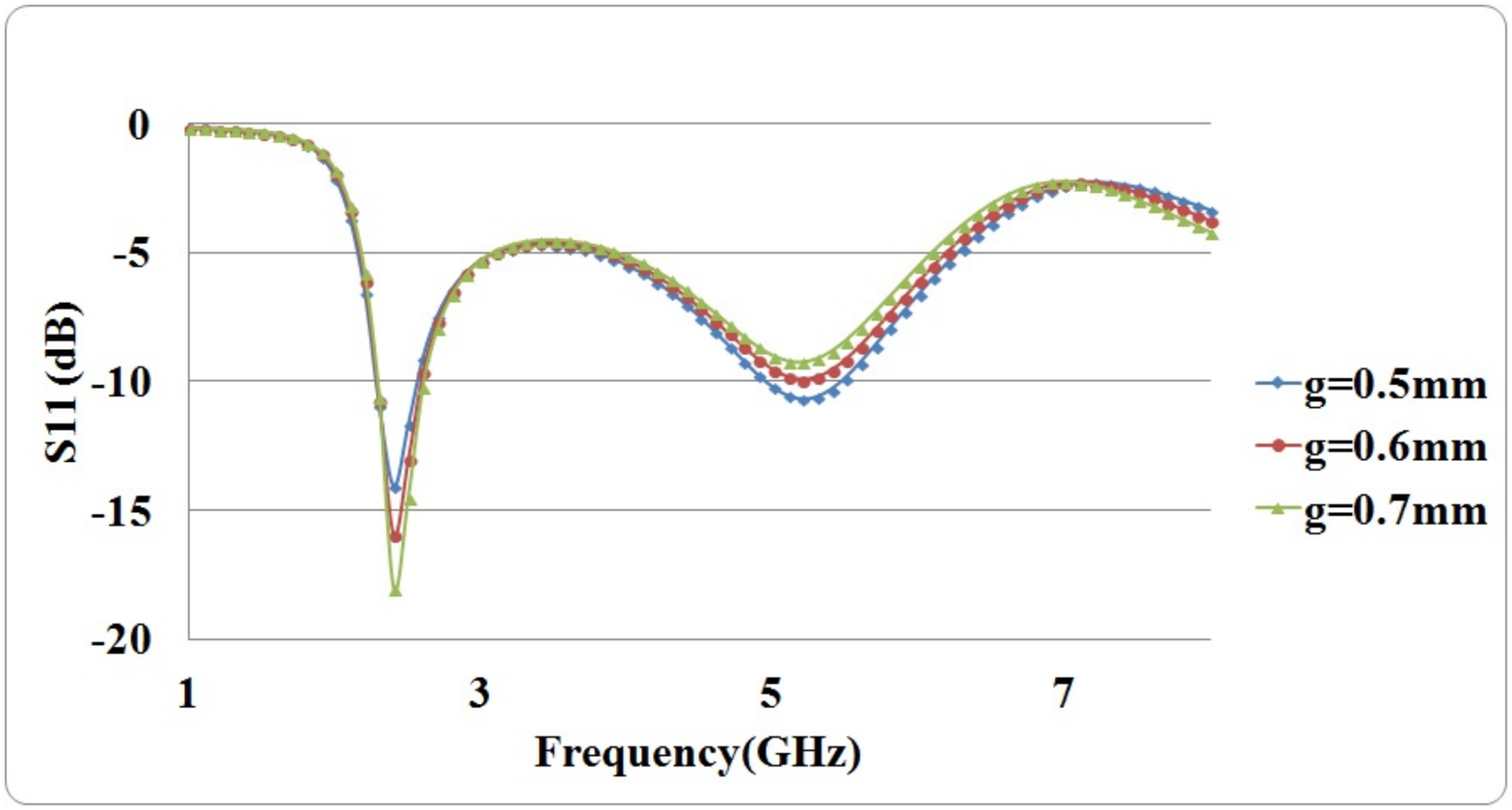}
\caption{Return Loss with variation in the coupling gap g}
\label{fig8}
\end{figure}

\begin{figure}[!]
\centering
\includegraphics[width=1\linewidth]{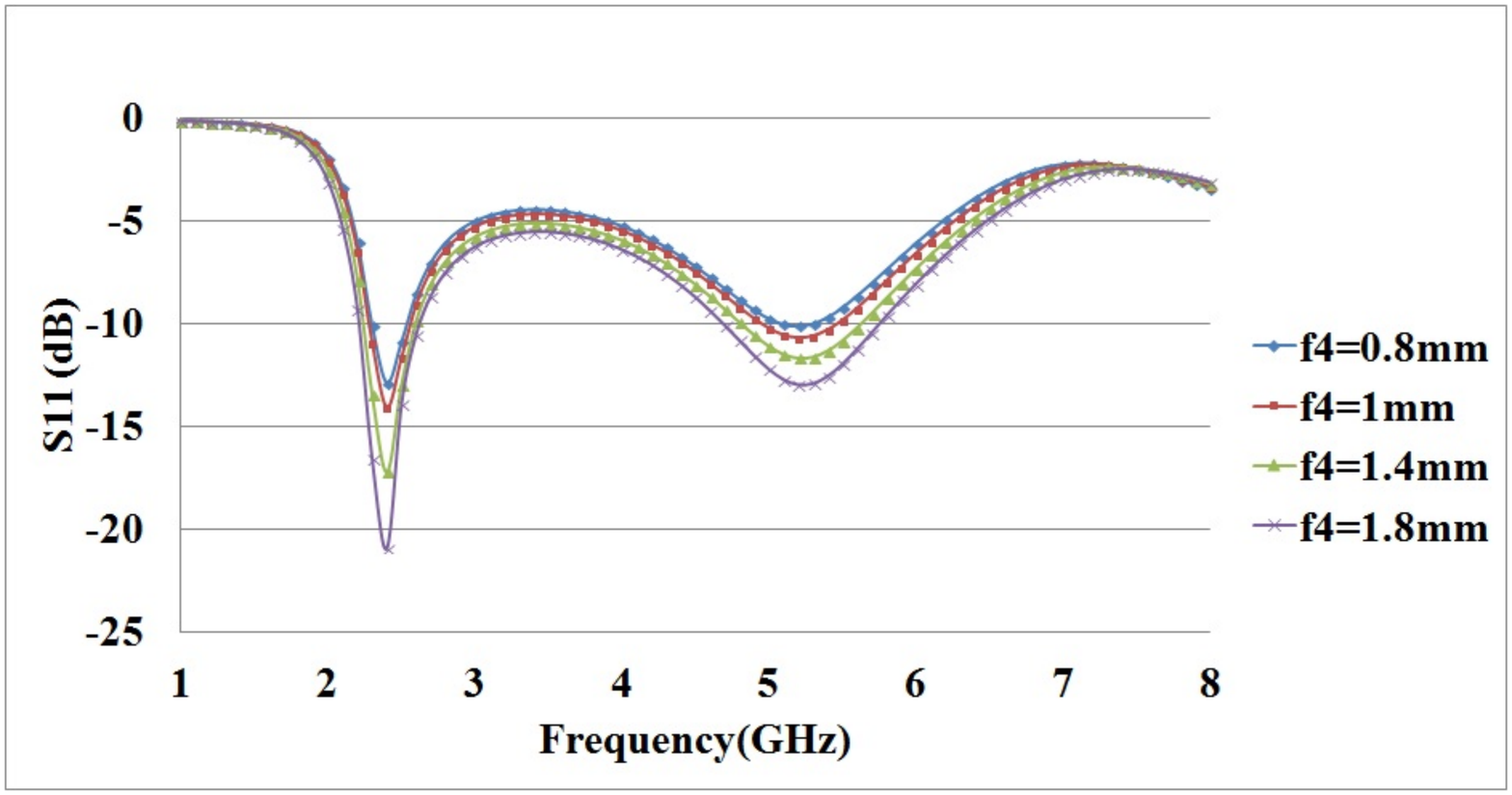}
\caption{Return Loss with variation in the value of f$4$}
\label{fig9}
\end{figure}

\begin{figure}[!]
\centering
\includegraphics[width=1\linewidth]{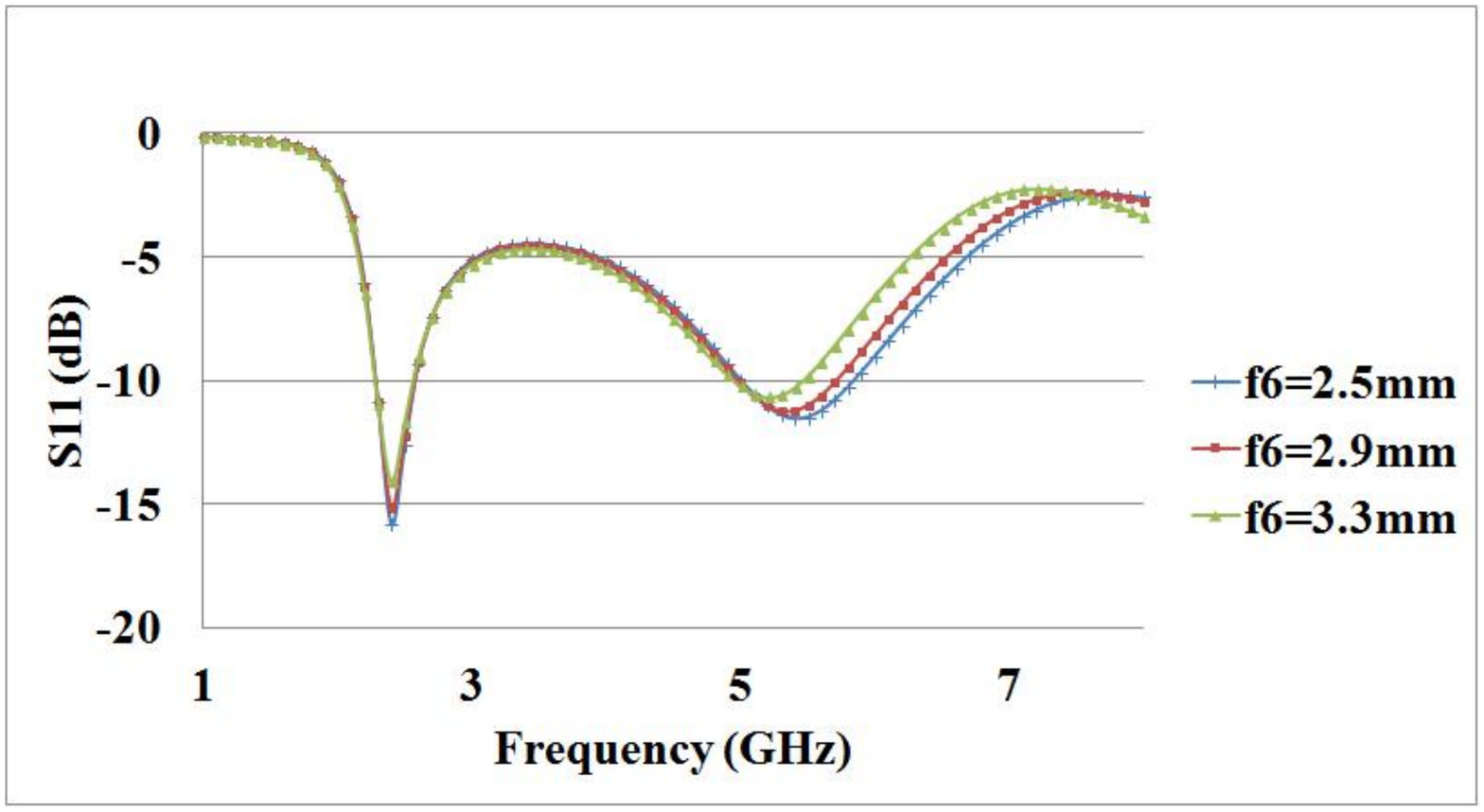}
\caption{Return Loss with variation in the value of f$6$}
\label{fig10}
\end{figure}

A detailed parametric study was performed to observe the effect of various dimensions on the two resonant frequencies. In order to study the resonating frequency of the two radiating elements, the antenna was simulated using Ansoft HFSS with the T-shape element alone and a single band was obtained at $6.5$ GHz. With both the elements present, a second reflection coefficient plot was obtained. The two plots have been shown in Fig.~\ref{fig5} and Fig.~\ref{fig3}. It can be clearly seen that when the F-element was added, a second resonant mode was generated at lower frequency of $2.4$ GHz while the higher frequency shifted to around $5.2$ GHz.\\
\indent Current distribution was also studied to gain further insight into the operation of the dual band antenna. The simulation results shows that at $2.4$ GHz, the current was mainly on the F-element which contributed to resonance. Similarly, at $5.2$ GHz, a large current was observed on the T-element which was responsible for that band. Fig.~\ref{fig6} shows the simulated current distribution at $2.4$ GHz and $5.2$ GHz.\\
\indent From computer simulation results, it was confirmed that the resonant frequency was sensitive to variation in the antenna parameters such as f$3$, f$4$, f$6$ and g. The value of f$3$ was varied and return loss plots as shown in Fig.~\ref{fig7} were observed. It can be clearly seen that as the value of f$3$ decreased, the resonant frequencies of both the upper and the lower band changed with the higher band frequency changing considerable. Also, the matching for the upper band improved significantly and the bandwidth also increased by a large amount. Hence, f$3$ can be used to achieve a coarse tuning of both the bands. Also, a resonant frequency of $5.8$ GHz was obtained while varying this length.\\ 
\indent Next, the gap g between the two elements was increased from $0.5$ mm to $0.6$ mm, $0.7$ mm. As expected, an upward shift was noticed as shown in Fig.~\ref{fig8} due to a decrease in the coupling between the two elements.\\
\indent The simulated S$11$ for variation in the value of f$4$ is shown in Fig.~\ref{fig9}. As the value of f$4$ increased, a good return loss was obtained while maintaining both the resonant frequencies constant. Thus, this dimension could be changed to get the required matching without disturbing the operating frequency.\\
\indent Finally, the value of f$6$ was decreased by $0.4$ mm and then by another $0.4$ mm. The corresponding return loss plot is shown in Fig.~\ref{fig10}. By changing the value of f$6$, the upper resonant frequency moved from $5.2$ GHz to $5.35$ GHz and finally $5.4$ GHz. The resonant frequency of the lower band, however, remained constant as f$6$ was varied. This proves that this dimension can be used to achieve a tuning of the upper resonant frequency while maintaining the lower operating frequency constant. Also, the impedance matching of both the bands improved.

\section{Design Methodology for other operating frequencies}
Based on the results of the parametric study, a generalized design methodology may be proposed for achieving a dual band performance at other frequencies.
\begin{enumerate}
\item In the first step, we set the values of t$2$ and t$4$, such that t$2+$t$4=\lambda_{g}/4$ and the ratio of t$2$ : t$4$ is the same as that in Table~\ref{table2}. $\lambda_{g}$ is the guide wavelength. By doing so, an operating frequency higher than the desired upper frequency is obtained. The symmetry should be maintained in the T-element and the F-element should not be present. 
\item Next, we insert the F-element and adjust the values of f$2$, f$3$ and f$6$ such that f$2+$f$3+$f$6=\lambda_{g}/4$. This may not exactly match the desired lower frequency. Also, the higher frequency may have shifted during this step. The ratio f$2$ : f$3$ : f$6$, the value of the gap `g' and the dimensions of the via must be maintained the same as the optimized values mentioned in Table~\ref{table2}.  
\item Since, the values of the operating frequencies obtained in Step $1$ and $2$ do not match the desired values, we use the result of the parametric study to achieve the required dual band operation. First, the value of the lower frequency should be adjusted by varying f$3$. The upper frequency will shift during this process.
\item Now, the value of f$6$ should be adjusted to tune the higher frequency to the desired value. When f$6$ is varied, the value of the lower frequency is maintained constant. Thus, we obtain the desired upper and lower desired resonant frequencies.
\item Finally, the value of f$4$ can be adjusted to obtaining a good impedance matching without disturbing the position of the two frequency bands. 
\end{enumerate}
Thus, by using this generalized approach, a dual band operation can also be obtained for other wireless systems in which the lower and higher operating frequencies lie in the ranges of $2.2$-$2.6$ GHz and $5.2$-$6.0$ GHz respectively. 
%

\begin{figure}[!]
  \centering
  \subfigure[]
  {
      \includegraphics[width=0.7\linewidth]{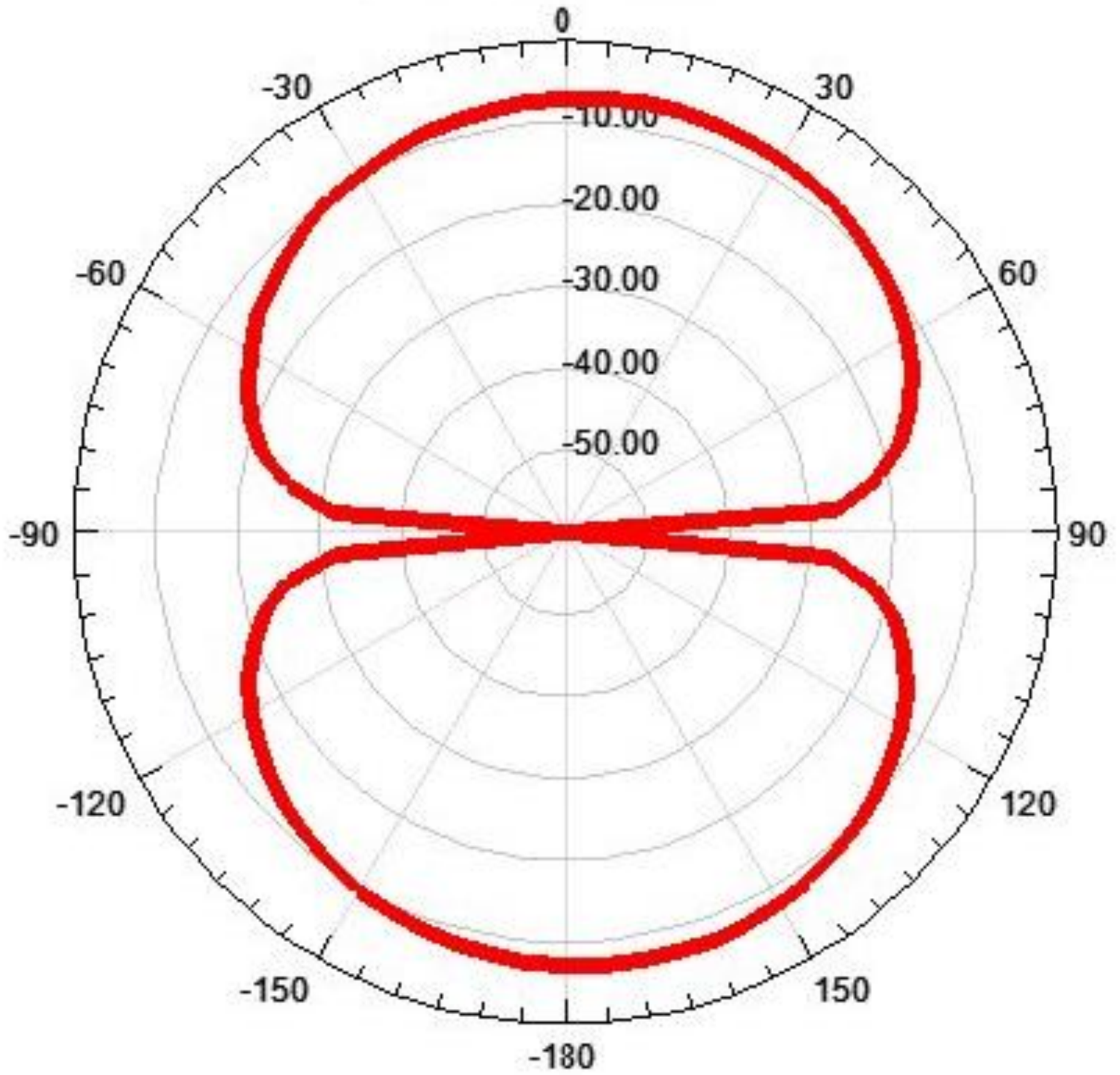}
      \label{ellipse}
  }
  \subfigure[]
  {
      \includegraphics[width=0.7\linewidth]{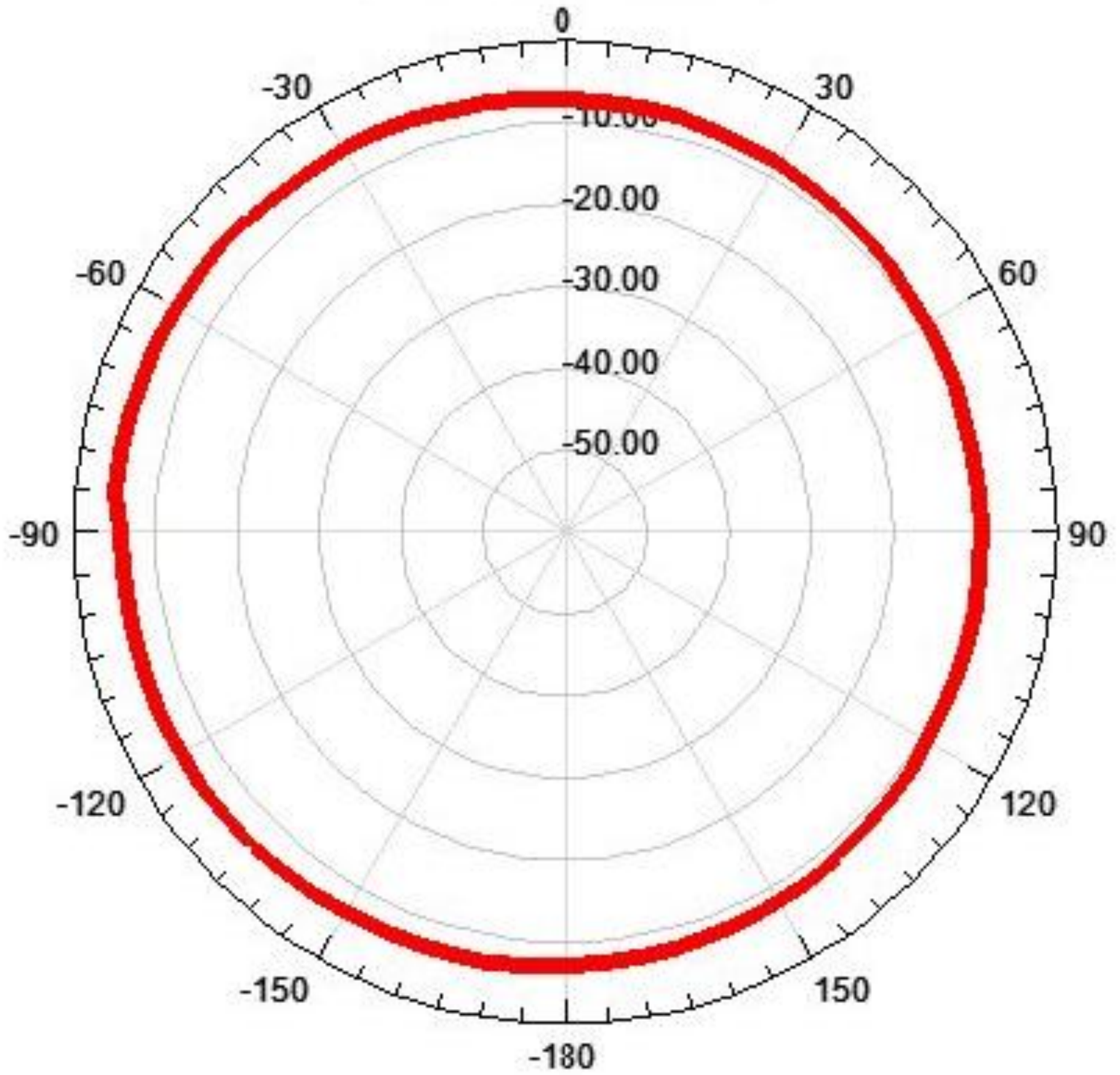}
      \label{circle}
  }
\\  
\subfigure[]
  {
      \includegraphics[width=0.7\linewidth]{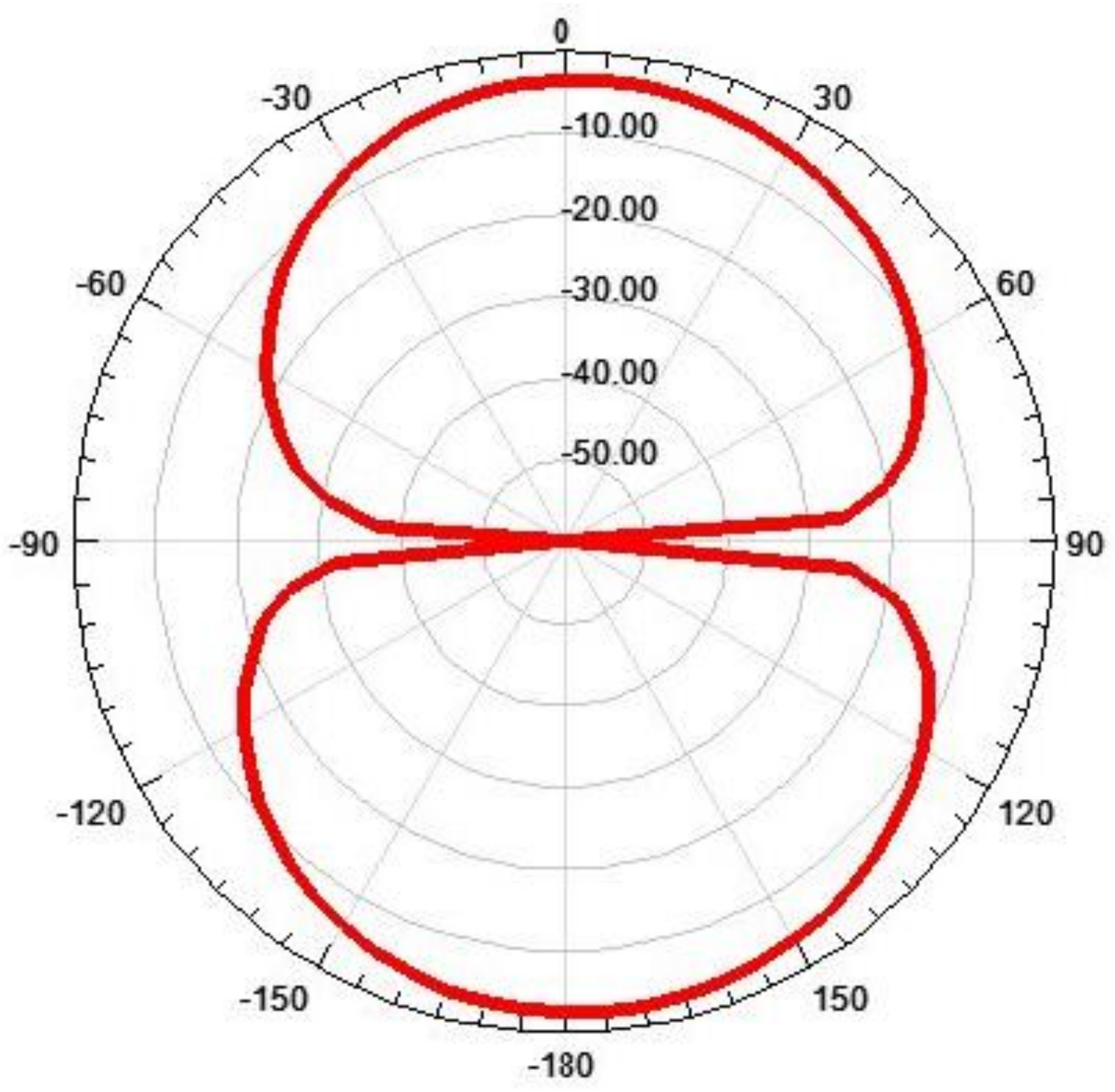}
      \label{drop}
  }
\\
\subfigure[]
  {
      \includegraphics[width=0.7\linewidth]{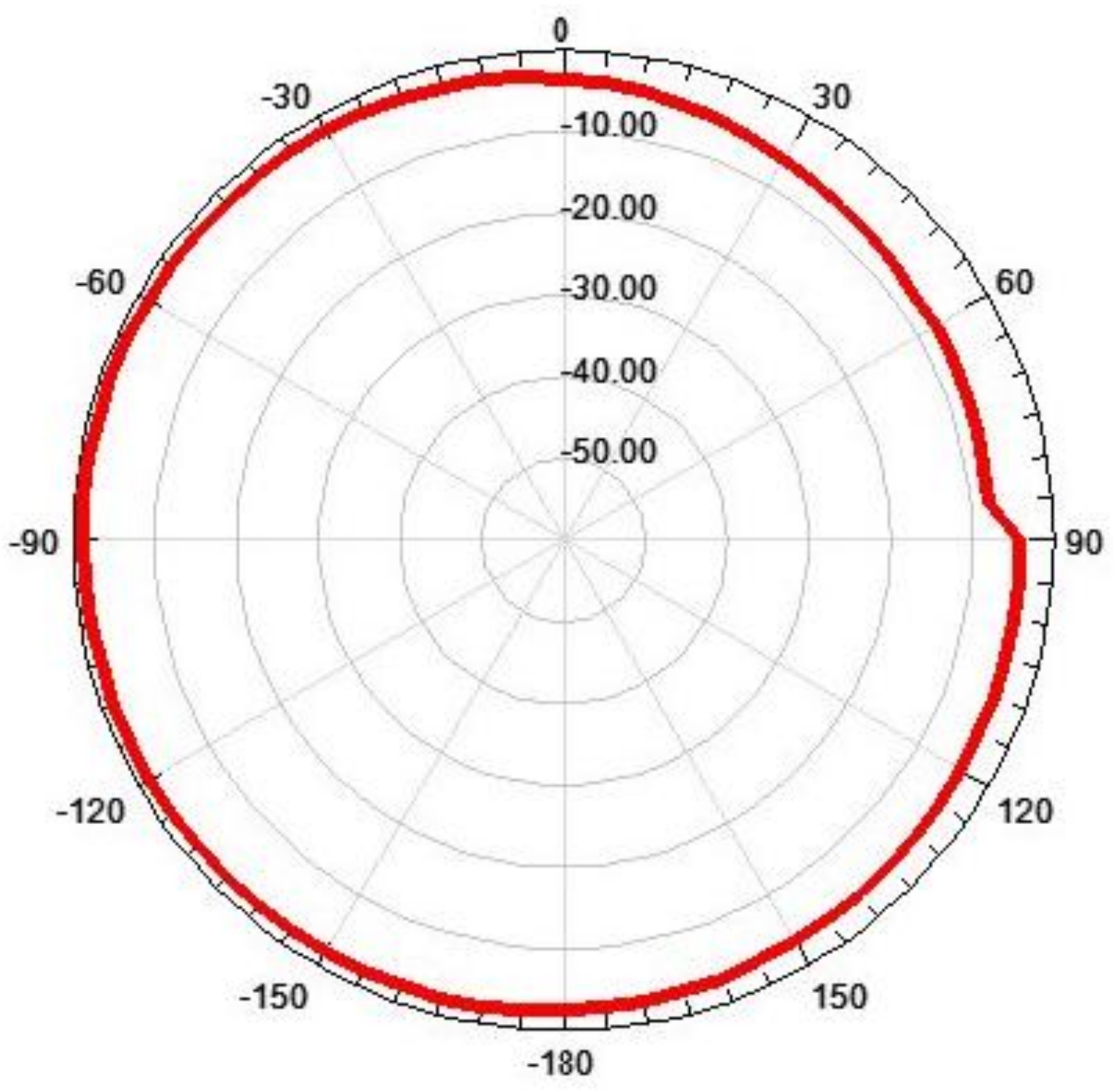}
      \label{drop}
  }
    
\caption{Simulated Radiation pattern: (a)E-plane pattern at 2.4 GHz  (b) H-plane pattern at 2.4 GHz (c) E-plane pattern at 5.2 GHz (d) H-plane pattern at 5.2 GHz}
  \label{rad}
\end{figure}

\section{Results and discussion}
The S$11$ and the radiation pattern simulation plots have been obtained with the help of Ansoft HFSS. The simulation and measured results of the return loss have been shown in Fig~\ref{fig3}. The bandwidths (S$11 < -10$ dB) for the lower band was from $2.27$ GHz to $2.58$ GHz and for the higher band was from $4.92$ GHz to $5.49$ GHz. Thus, the lower and upper frequency band of WLAN systems are satisfactorily covered.\\
\indent The simulated radiation pattern at $2.4$ GHz and $5.2$ GHz are shown in Fig~\ref{rad}. It can be clearly observed that the antenna has an omnidirectional radiation pattern at both the lower as well as the higher frequencies. This is one of the important properties desired of a antenna for WLAN systems. The  radiation efficiency at $2.4$ GHz and $5.2$ GHz are $89$\% and $87$\%.\\
\indent The compact size of the antenna enable the designer to save a lot of space on the device with which it is to be integrated. The designer may use the saved space to increase the number of other circuit elements and add additional features to the device. Also, due to its small size, its fabrication cost gets considerably reduced. Additionally, the antenna can be tuned to other frequencies for dual band operation by merely changing the dimensions.

\section{Conclusion}
A novel and compact planar monopole dual band antenna for $2.4$/$5.2$/$5.8$ GHz WLAN applications has been designed and studied in this paper. The antenna has a compact radiator of area $11\times6.5$ mm$^{2}$ comprising of a T-element and a F-element resonating at $5.2$/$5.8$ and $2.4$ GHz respectively. By conducting a thorough parametric study, it has been found that these two frequency bands can be tuned independently by varying certain dimensions of the antenna. Thus, the antenna can be designed to obtain a dual band performance at other frequencies by merely changing the key dimensions. The plots of reflection coefficient S$11$ and radiation pattern for the two frequencies have been studied. These results have shown that the design is a very promising candidate for WLAN applications.

\ifCLASSOPTIONcaptionsoff
  \newpage
\fi

\bibliographystyle{unsrt}
\bibliography{paper}

\end{document}